\documentclass[twocolumn,showpacs,superscriptaddress,amsmath,amssymb,nofootinbib,aps, prd]{revtex4-2}
\usepackage{graphicx}
\usepackage{overpic}
\usepackage{epsfig}
\usepackage{dcolumn}
\usepackage{multirow}
\usepackage{booktabs}
\usepackage{appendix}
\usepackage{bm}
\usepackage{threeparttable}
\usepackage{lineno}
\usepackage[colorlinks,linkcolor=blue,anchorcolor=blue,citecolor=blue]{hyperref}
\usepackage{epstopdf}
\usepackage{overpic}
\usepackage{color}
\usepackage{float}
\usepackage{subfigure}
\usepackage[capitalize]{cleveref}
\makeatletter
\renewcommand{\maketag@@@}[1]{\hbox{\m@th\normalsize\normalfont#1}}%
\makeatother

\def\pip{\pi^{+}}
\def\pim{\pi^{-}}
\def\piz{\pi^{0}}
\def\ee{e^{+}e^{-}}
\def\dedx{\mathrm{d}E/\mathrm{d}x}

\def \Ks {K_{S}^{0}}

\def \gevcc{\mbox{GeV/$c^2$}}
\def \mev  {\mbox{MeV}}

\def \mpipi {M_{\pi^{+}\pi^{-}}}
\def \miss2{M_{\rm miss}^{2}}

\def \romanOne   {\uppercase\expandafter{\romannumeral1}}
\def \romanTwo   {\uppercase\expandafter{\romannumeral2}}
\def \romanThree {\uppercase\expandafter{\romannumeral3}}
\def \romanFour  {\uppercase\expandafter{\romannumeral4}}
\def \romanFive  {\uppercase\expandafter{\romannumeral5}}
\def \romanSix   {\uppercase\expandafter{\romannumeral6}}
\def \romanSeven {\uppercase\expandafter{\romannumeral7}}
\def \romanEight {\uppercase\expandafter{\romannumeral8}}
\def \romanNine {\uppercase\expandafter{\romannumeral9}}

\newcommand{\lamcplamcm}{\Lambda_{c}^{+}\bar{\Lambda}_{c}^{-}}
\newcommand{\lambdacp}{\Lambda_{c}^{+}}
\newcommand{\lambdacm}{\bar{\Lambda}_{c}^{-}}

\newcommand{\sigmode}[1]{
	\ifnum#1=1
	\lambdacp \rightarrow n K_{S}^{0} \pi^{+}
	\else
	\ifnum#1=2
	\lambdacp \rightarrow n K_{S}^{0} K^{+}
	\fi
	\fi
}

\renewcommand{\figurename}{Figure}

\begin{document}
\title{\boldmath Measurement of Branching Fractions for $\sigmode{1}$ and $\sigmode{2}$}
\author{
\small
M.~Ablikim$^{1}$, M.~N.~Achasov$^{5,b}$, P.~Adlarson$^{75}$, X.~C.~Ai$^{81}$, R.~Aliberti$^{36}$, A.~Amoroso$^{74A,74C}$, M.~R.~An$^{40}$, Q.~An$^{71,58}$, Y.~Bai$^{57}$, O.~Bakina$^{37}$
% ...
\\
\vspace{0.2cm}
(BESIII Collaboration)\\
\vspace{0.2cm} {\it
$^{1}$ Institute of High Energy Physics, Beijing 100049, People's Republic of China\\
% ...
$^{5}$ Budker Institute of Nuclear Physics SB RAS (BINP), Novosibirsk 630090, Russia\\
% ...
$^{36}$ Johannes Gutenberg University of Mainz, Johann-Joachim-Becher-Weg 45, D-55099 Mainz, Germany\\
$^{37}$ Joint Institute for Nuclear Research, 141980 Dubna, Moscow region, Russia\\
% ...
$^{40}$ Liaoning Normal University, Dalian 116029, People's Republic of China\\
% ...
$^{57}$ Southeast University, Nanjing 211100, People's Republic of China\\
$^{58}$ State Key Laboratory of Particle Detection and Electronics, Beijing 100049, Hefei 230026, People's Republic of China\\
% ...
$^{71}$ University of Science and Technology of China, Hefei 230026, People's Republic of China\\
% ...
$^{74}$ University of Turin and INFN, (A)University of Turin, I-10125, Turin, Italy; (B)University of Eastern Piedmont, I-15121, Alessandria, Italy; (C)INFN, I-10125, Turin, Italy\\
$^{75}$ Uppsala University, Box 516, SE-75120 Uppsala, Sweden\\
% ...
$^{81}$ Zhengzhou University, Zhengzhou 450001, People's Republic of China\\
\vspace{0.2cm}
% ...
$^{b}$ Also at the Novosibirsk State University, Novosibirsk, 630090, Russia\\
% ...
}\vspace{0.4cm}
}
\date{\today}
\begin{abstract}
Based on 4.5 fb$^{-1}$ of $\ee$ collision data accumulated at center-of-mass energies between $4599.53\,\mev$ and $4698.82\,\mev$ with the BESIII detector, we measure the absolute branching fraction of the Cabibbo-favored decay $\sigmode{1}$ with the precision improved by a factor of 2.8 and report the first evidence for the singly-Cabibbo-suppressed decay $\sigmode{2}$. 
The branching fractions for $\sigmode{1}$ and $\sigmode{2}$ are determined to be $(1.86\pm0.08\pm0.04)\times10^{-2}$ and $\left(3.9^{+1.7}_{-1.4}\pm0.3\right)\times10^{-4}$, respectively, where the first uncertainties are statistical and the second ones are systematic.
\end{abstract}
\maketitle

\setrunninglinenumbers


\section{\boldmath Introduction}

\begin{figure}[tbp]\centering
	\hspace{-3mm}
	\subfigure[]{
		\label{fig:nkspiFeynmanA}
		\includegraphics[width=0.20\textwidth]{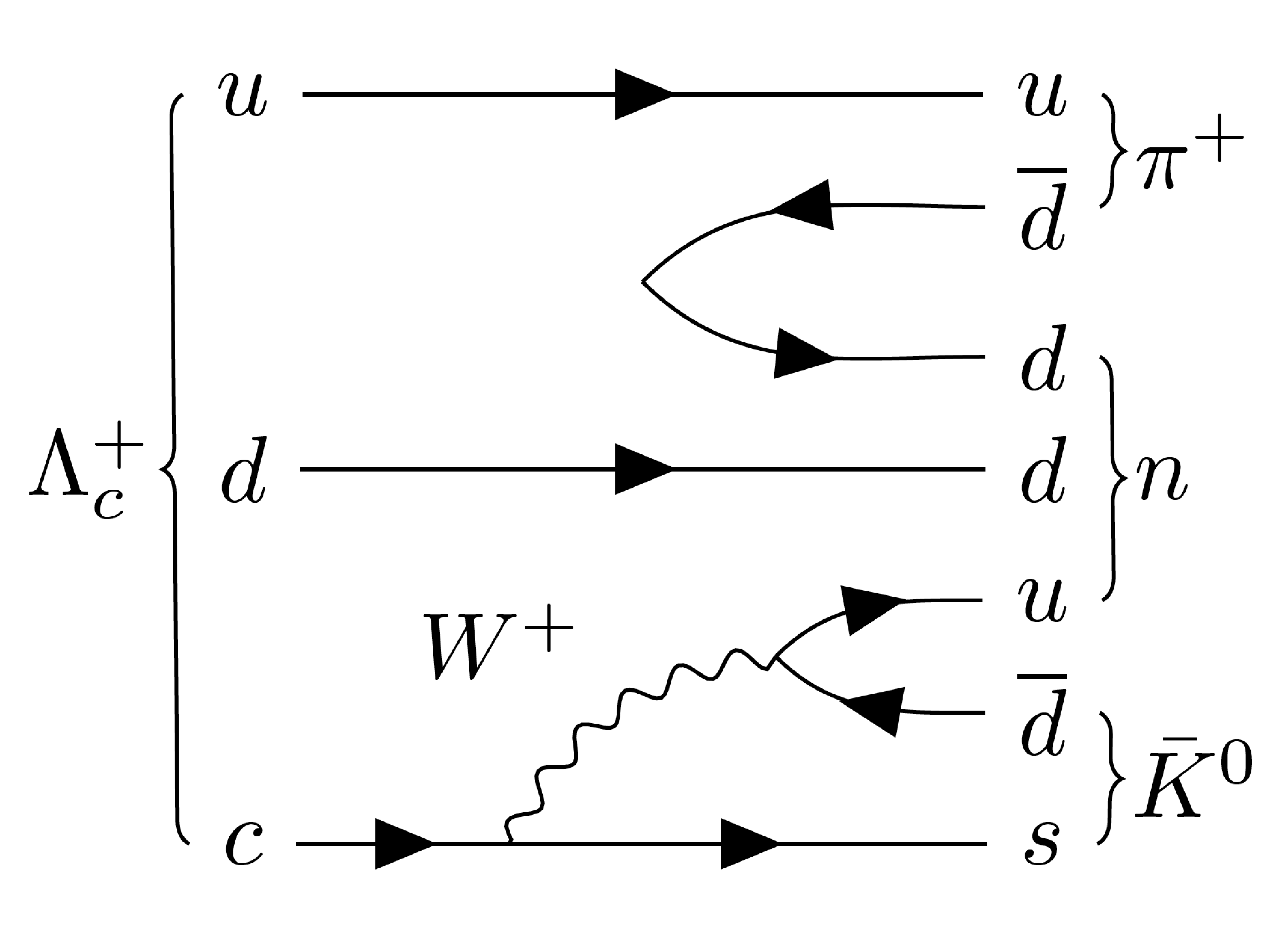}
	}
	\hspace{0mm}
	\subfigure[]{
		\label{fig:nkspiFeynmanB}
		\includegraphics[width=0.20\textwidth]{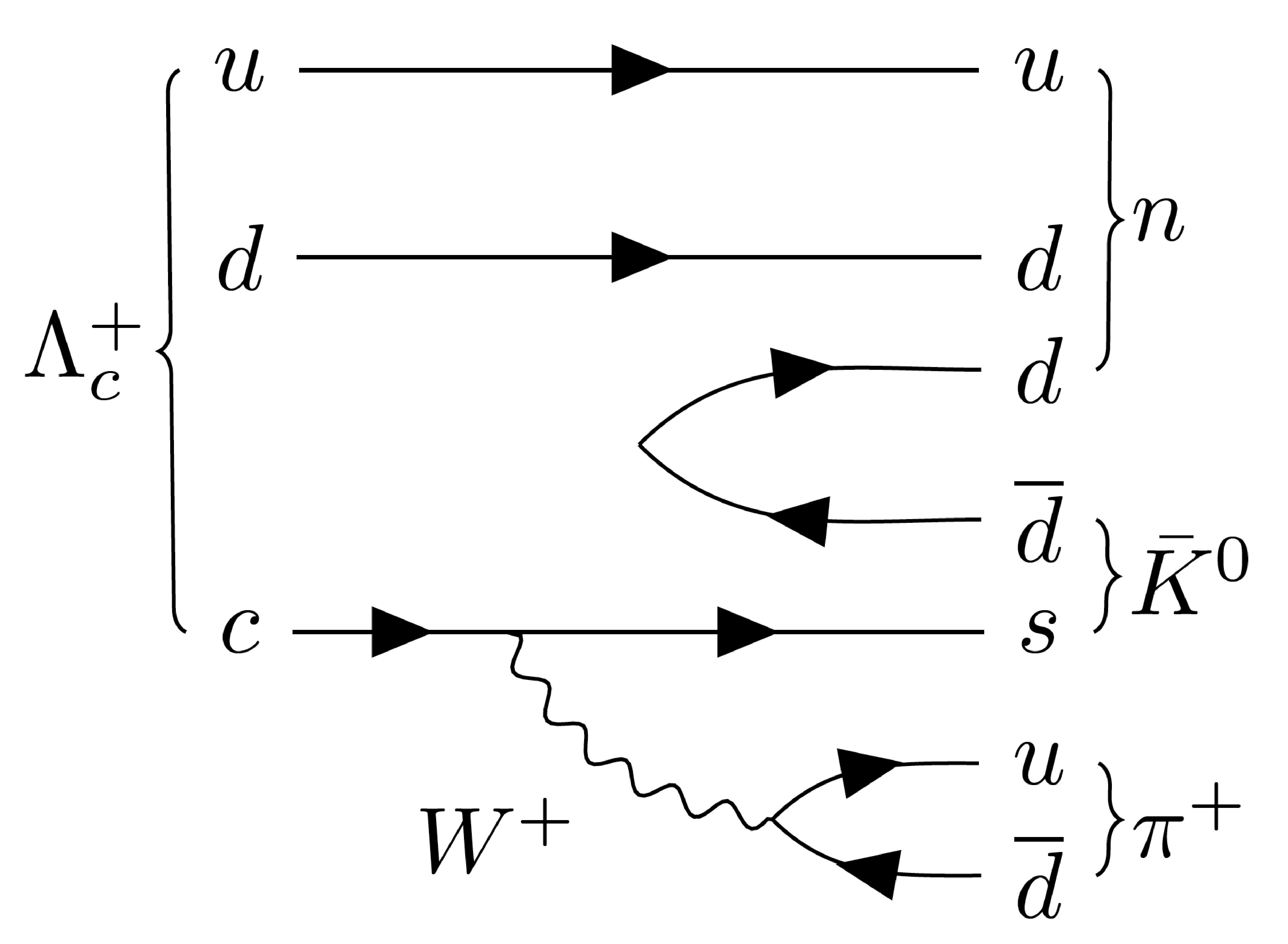}
	}
	\hspace{-15mm}
	\subfigure[]{
		\label{fig:nkspiFeynmanC}
		\includegraphics[width=0.20\textwidth]{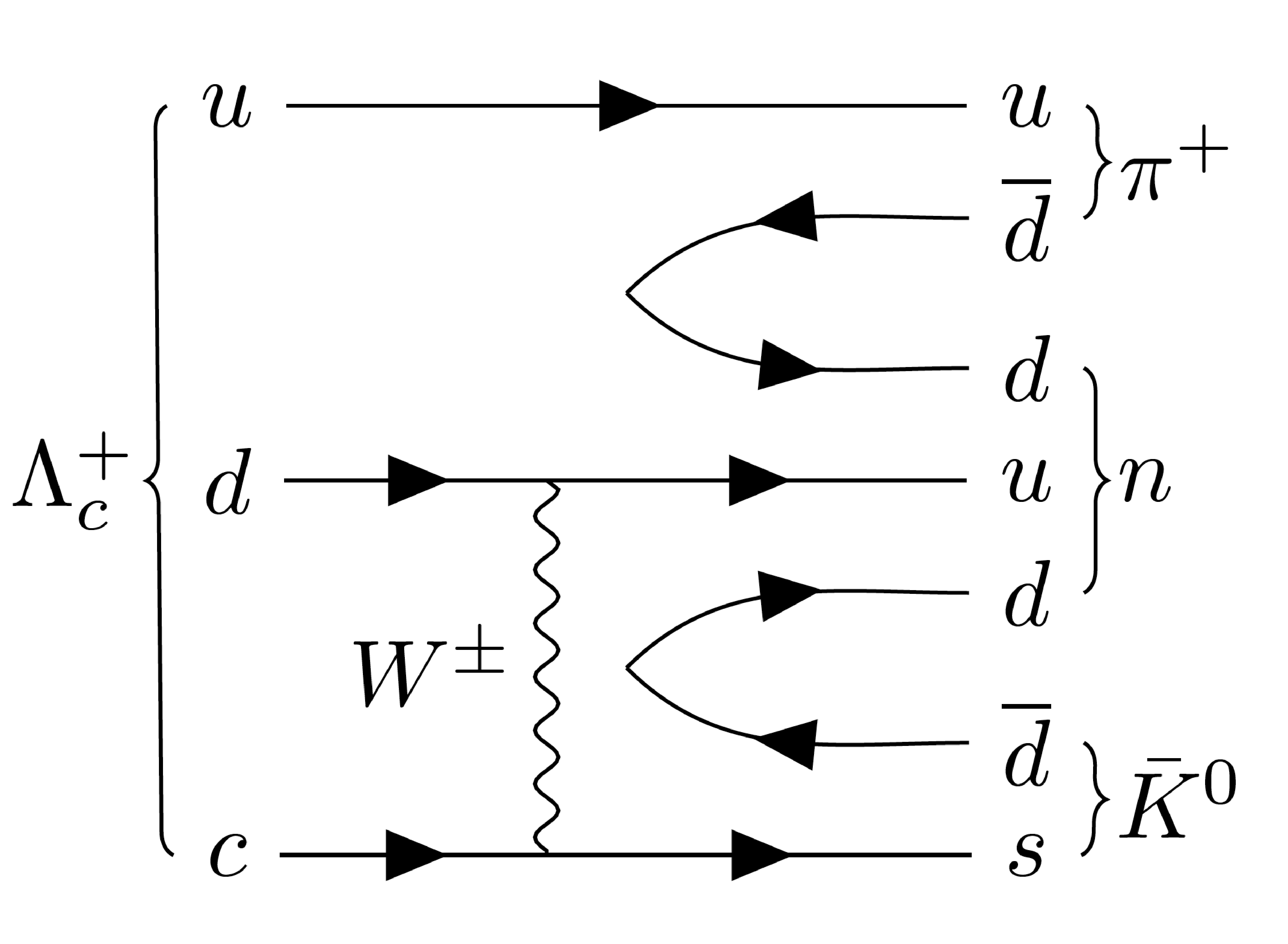}
	}
	\subfigure[]{
		\label{fig:nkskFeynmanA}
		\includegraphics[width=0.20\textwidth]{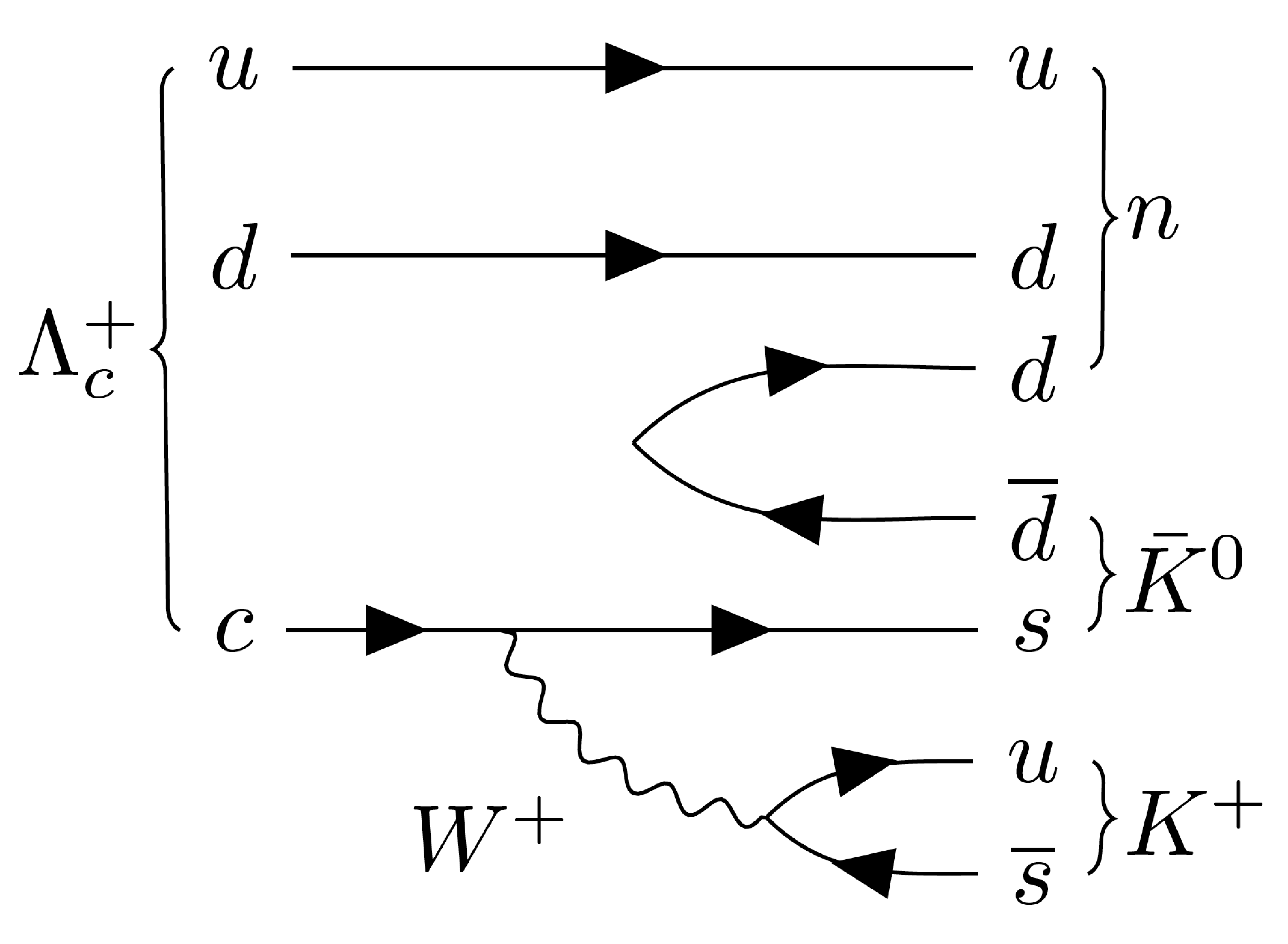}
	}
	\hspace{-15mm}
	\subfigure[]{
		\label{fig:nkskFeynmanC}
		\includegraphics[width=0.20\textwidth]{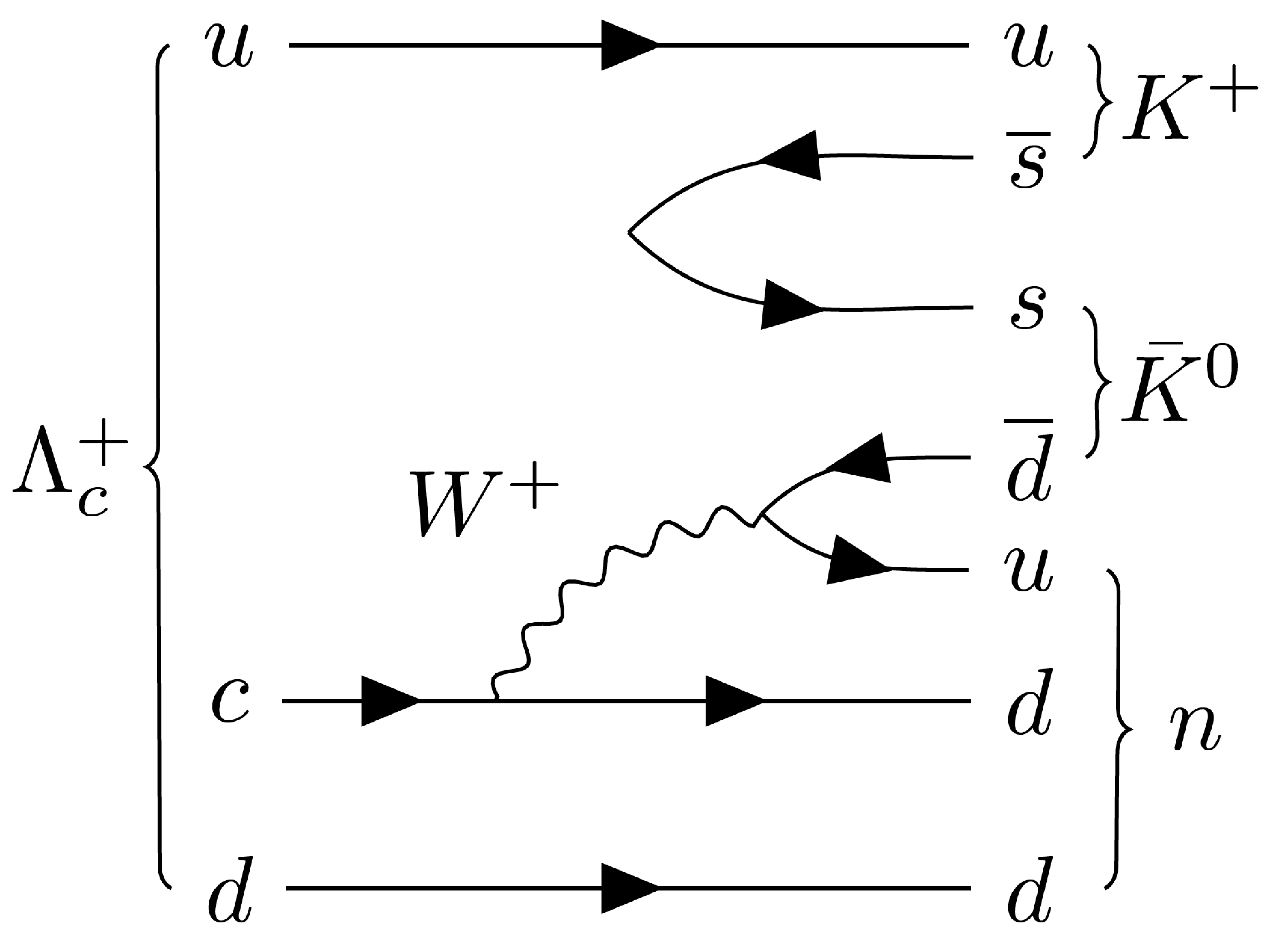}
	}
	\caption{Topological diagrams for (a, b, c) $\Lambda_{c}^{+}\to n\bar{K}^{0}\pi^{+}$ and (d, e) $\Lambda_{c}^{+}\to n\bar{K}^{0}K^{+}$.}
	\label{fig:Feynman}
\end{figure}

Studies of weak decays of charmed baryons provide crucial information on the dynamics of strong and weak interactions in charm physics.
Theoretical predictions for $\lambdacp$ decays are difficult. 
Decay amplitudes of charmed hadrons are split into two parts, factorizable and non-factorizable~\cite{Chau:1986jb, Chau:1995gk}. 
Both external and internal $W$-emission diagrams are mainly factorizable. Inner $W$-emission and $W$-exchange diagrams are non-factorizable.
For internal $W$-emission diagram, the quark produced by the $W$ emission forms part of a meson, as shown in Fig.~\ref{fig:nkspiFeynmanA}, while for the inner $W$-emission diagram, that quark forms part of a baryon \cite{Cheng:1993gf,CHENG2022324}, as shown in Fig.~\ref{fig:nkskFeynmanC}.
Unlike charmed mesons, $W$-exchange diagram, manifested as a baryon pole diagram, is no longer subject to helicity and color suppression, which makes theoretical calculations more complex. 
There has been much progress in the study of two-body decays of $\lambdacp$ in both theory and experiment~\cite{Cheng:2015iom,CHENG2022324}. However, the dynamics of three-body decays is more complicated due to the contributions of intermediate resonances and theoretical work on three-body decays is insufficient.

According to Ref.~\cite{Lu:2016ogy}, two isospin amplitudes $I^{(0)}$ and $I^{(1)}$ for $\sigmode{1}$ are defined as $N\bar{K}$ isospin singlet and isospin triplet. 
Based on isospin symmetry, the ratio $R$ between the moduli of the two isospin amplitudes is
\begin{footnotesize}
\begin{equation}
	\label{eq:isospinRatio}
	\frac{|I^{(1)}|}{|I^{(0)}|}\\=\sqrt{\frac{\mathcal{B}(pK^{-}\pip)+\mathcal{B}(n\bar{K}^{0}\pip)-\mathcal{B}(p\bar{K}^{0}\pi^{0})}{2\mathcal{B}(p\bar{K}^{0}\pi^{0})}},
\end{equation}
\end{footnotesize}
and their relative strong phase
\begin{scriptsize}
\begin{equation}
	\label{eq:isospinStrongPhase}
	\cos\delta = \frac{\mathcal{B}(n\bar{K}^{0}\pip)-\mathcal{B}(pK^{-}\pip)}{2\sqrt{\mathcal{B}(p\bar{K}^{0}\pi^{0})(\mathcal{B}(pK^{-}\pip)+\mathcal{B}(n\bar{K}^{0}\pip)-\mathcal{B}(p\bar{K}^{0}\pi^{0}))}}.
\end{equation}
\end{scriptsize}
Therefore, $R$ and $\cos\delta$ can be extracted using the measured branching fractions (BFs) of $pK^{-}\pip$, $p\bar{K}^{0}\pi^{0}$ and $n\bar{K}^{0}\pip$. 

The $\Lambda_{c}^{+}\to n\bar{K}^{0}\pi^{+}$ decay is one of the significant decays of the $\Lambda_{c}^{+}$ involving a neutron. 
Figures~\labelcref{fig:nkspiFeynmanA}, \labelcref{fig:nkspiFeynmanB}, and \labelcref{fig:nkspiFeynmanC} show the leading-order topological diagrams for $\Lambda_{c}^{+}\to n\bar{K}^{0}\pi^{+}$, which proceed via internal W-emission, external W-emission, and W-exchange, respectively. 
Hence, $\Lambda_{c}^{+}\to n\bar{K}^{0}\pi^{+}$ decay is dominated by the weak transition $c\to su\bar{d}$. 
The external $W$-emission diagram, as shown in Fig.{~\ref{fig:nkspiFeynmanB}}, where a $\pip$ is emitted and the $N\bar{K}$ forms an isospin singlet, is dominated by $I^{(0)}$. If factorization works, non-factorizable components, such as Fig.{~\ref{fig:nkspiFeynmanC}}, contribute far less than external and internal $W$-emission diagrams, so the amplitude of $\sigmode{1}$ is dominated by $I^{(0)}$, and the two independent isospin amplitudes are real with vanishing phases at leading order. The measured $R$ can be used to validate the factorization scheme in $\lambdacp$ decays, and the measured $\cos\delta$ provides essential input for the analysis of hadronic decays into other baryons and testing isospin symmetry. In 2014, BESIII measured the BF of $\lambdacp\to n\Ks\pip$ for the first time to be $(1.82\pm0.25)\%$ \cite{BESIII:2016yrc}. Combining with the known BFs of $\Lambda_c^+\to p\bar{K}^{0}\pi^{0}$ and $\Lambda_c^+\to pK^{-}\pip$~\cite{pdg2022}, $R$ and $\cos\delta$ are evaluated to be $1.14\pm0.11$ and $-0.24\pm0.08$, respectively.

The $\Lambda_{c}^{+}\to n\bar{K}^{0}K^{+}$ decay is a singly-Cabibbo-suppressed process. Figures~\labelcref{fig:nkskFeynmanA,fig:nkskFeynmanC} show the leading-order topological diagrams for $\Lambda^+_c\to n\bar K^0K^+$, which proceed via external W-emission and inner W-emission, respectively. 
Hence, the $\Lambda_{c}^{+}\to n\bar{K}^{0}K^{+}$ decay is dominated by two weak transitions, $c\to s\bar{s}u$ and $c\to d\bar{d}u$. 
Therefore, we cannot define physical quantities based on isospin symmetry for $\lambdacp\to n\bar{K}^{0}K^{+}$ and its isospin partners $\lambdacp\to pK^{+}K^{-}$ and $\lambdacp\to pK^{0}\bar{K}^{0}$. 
Instead, the measurement of the BF of $\sigmode{2}$ can help us to understand the non-factorizable contribution of $\lambdacp$ decays. However, there is no experimental measurement of $\sigmode{2}$ available yet.



In this paper, we report the measurement of branching fraction of $\sigmode{1}$ with an improved precision and the first evidence of $\sigmode{2}$, using 4.5 $\rm fb^{-1}$ of $e^{+}e^{-}$ collision data collected at center-of-mass (c.m.) energies between $4599.53\,\mev$ and $4698.82\,\mev$ with the BESIII detector. 
Since these energy points are just above the $\lambdacp\lambdacm$ pair production threshold, the $\lambdacp\lambdacm$ pairs are produced cleanly without additional fragmentation hadrons, which makes it feasible to apply the double-tag (DT) method~\cite{Li:2021iwf} and reconstruct the neutron with a missing-mass technique. 
The $\lambdacm$, denoted as single-tag (ST) candidate, is reconstructed using eleven exclusive hadronic decay modes, as listed in \tablename~\ref{tab:N_ST}. The $\lambdacp$ is reconstructed in the system recoiling against the ST candidate, and an event containing an ST $\bar{\Lambda}^-_c$ and a signal $\Lambda^+_c$ is denoted as the DT candidate. 
Charge conjugation is always implied throughout this paper.


\section{\boldmath BESIII Experiment and Monte Carlo Simulation}
The BESIII detector~\cite{Ablikim:2009aa} records symmetric $e^+e^-$ collisions provided by the BEPCII storage ring~\cite{Yu:2016cof} in the c.m.~energy range from 2.0 to 4.95~GeV, with a peak luminosity of $1.0 \times 10^{33}\;\rm{cm^{-2}}\rm{s}^{-1}$ achieved at a c.m.~energy of $\sqrt{s} = 3.77\;\rm{GeV}$. 
BESIII has collected large data samples in this energy region~\cite{BESIII:2020nme}.
The cylindrical core of the BESIII detector covers 93\% of the full solid angle and comprises a helium-based multilayer drift chamber (MDC), a plastic scintillator time-of-flight system (TOF), and a CsI(Tl) electromagnetic calorimeter (EMC), which are all enclosed in a superconducting solenoidal magnet providing a 1.0 T magnetic field. The solenoid is supported by an octagonal flux-return yoke which is segmented into layers and instrumented with resistive plate counters modules for muon identification. The charged-particle momentum resolution at $1\;\rm{GeV}/c$ is 0.5\%, and ionization energy loss $\dedx$  resolution is 6\% for electrons from Bhabha scattering. The EMC measures photon energies with a resolution of 2.5\% (5\%) at 1 GeV in the barrel (end cap) region. The time resolution in the TOF barrel region is 68 ps, while that in the end cap region was 110 ps. The end cap TOF system was upgraded in 2015 using multi-gap resistive plate chamber technology, providing a time resolution of 60 ps~\cite{li:TOF1, guo:TOF2, Cao:2020ibk}. About 85\% of the $\lambdacp\lambdacm$ pairs are produced in data taken after this upgrade. More detailed descriptions can be found in Refs. ~\cite{Ablikim:2009aa, Yu:2016cof}.

Simulated data samples are produced with a {\sc geant4}-based~\cite{geant4} Monte-Carlo (MC) package, which includes the geometric description of the BESIII detector~\cite{GDMLMethod,BesGDML,Huang:2022wuo} and the time-dependent detector response. The simulation models the beam-energy spread and initial-state radiation (ISR) in the $\ee$ annihilations with the generator {\sc kkmc}~\cite{kkmc}. Final-state radiation from charged final-state particles is incorporated using {\sc photos}~\cite{photos} package.

The ``inclusive MC sample'' includes the production of $\lamcplamcm$ pairs, open-charmed mesons, ISR production of vector charmonium(-like) states, and continuum processes which are incorporated in {\sc kkmc}~\cite{kkmc, kkmc2}. All the known decay modes are modeled with {\sc evtgen}~\cite{evtgen, besevtgen} using the BFs taken from the PDG~\cite{pdg2022}. The remaining unknown charmonium decays are modeled with {\sc lundcharm}~\cite{lundcharm, Yang:2014vra}. The inclusive MC sample is used to determine the ST efficiencies and estimate backgrounds. 
The ``signal MC sample'' denotes the exclusive processes where $\lambdacm$ decays to eleven ST modes and $\lambdacp$ decays to $\sigmode{1}$ or $\sigmode{2}$, with $\Ks\to\pip\pim$. The signal MC samples are used to evaluate the DT efficiencies and extract the signal shapes. The $\sigmode{1}$ and $\sigmode{2}$ signal MC samples are simulated with a phase space (PHSP) model, where the events are evenly distributed in PHSP. For $\sigmode{1}$, the two-body invariant mass distributions have been weighted to match those of data, as detailed in Sec.~\ref{sec:bf}. 
The ``exclusive background MC sample'' denotes the exclusive processes where $\lambdacm$ decays to eleven ST modes and $\lambdacp$ decays to $n\pip\pim\pip$, $\Sigma^{+}\pim\pip$ and $\Sigma^{-}\pip\pip$. The MC samples for the dominant backgrounds are utilized to estimate the contamination rates and extract peaking background shapes.

\section{\boldmath Event Selection}
\label{sec:selection}
The selection criteria for ST candidates are the same as Ref.~\cite{BESIII:2022xne}. 
The ST $\lambdacm$ baryons are identified with beam-constrained mass $M_{\rm BC}\equiv\sqrt{E_{\rm beam}^{2}/c^{4}-p^{2}/c^{2}}$, where $E_{\rm beam}$ is the beam energy and $p$ is the measured momentum of $\lambdacm$ in the c.m.~system of $e^+e^-$ collision. 
The signal and sideband regions for ST candidates are chosen as $(2.280, 2.296)\,\gevcc$ and $(2.250, 2.270)\,\gevcc$, respectively.
Candidates falling in the signal region are retained for further signal-side reconstruction, and those falling in the sideband region are used to estimate background contributions.

Two signal channels are reconstructed through the decays $\sigmode{1}$ and $\sigmode{2}$ with $\Ks\to\pip\pim$, recoiling against the ST candidates.
Charged tracks detected in the MDC are required to be within a polar angle ($\theta$) range of $|\rm{cos\theta}|<0.93$, where $\theta$ is defined with respect to the $z$-axis, which is the symmetry axis of the MDC. 
The $\Ks$ candidate is reconstructed from two oppositely charged tracks satisfying $|V_{z}|<20\,\mathrm{cm}$, where $V_{z}$ denotes the closest distance from the track itself to the $\ee$ interaction point (IP) along $z$-axis. No distance constraint in $xy$ plane is required.
A loose particle identification (PID)~\cite{Asner:2008nq} requirement is imposed on the two charged tracks. 
The loose PID procedure uses information from either the time of flight in the TOF or the $\dedx$ in the MDC to calculate $\chi^2_h$ ($h=\pi, K$) for each hadron $h$ hypothesis. 
Charged tracks from the $\Ks$ are identified as pions when either $\chi^2_{\pi}$ or $\chi^2_{K}$ is less than 4. 
The pions are constrained to originate from a common vertex. 
The decay length of the $\Ks$ candidate is required to be greater than twice the vertex resolution away from the IP, $i.e.$, $L/\sigma_{L}>2$, where $L$ and $\sigma_{L}$ denote the three-dimensional (3D) decay length and its uncertainty, respectively. If there are multiple $\Ks$ candidates, the one with the largest $L/\sigma_{L}$ is kept. 

Apart from the $\Ks$ candidate, we require one additional charged track. 
The remaining charged pion (kaon) is further required to satisfy $|V_{z}|<10\,\mathrm{cm}$ and $V_r<1\,\mathrm{cm}$, where $V_r$ denotes the distance to the IP in $xy$ plane, and they must also pass the requirements on the loose PID $\chi^{2}_{\pi, K}$ given above. 
The PID likelihood $\mathcal{L}(h)$ ($h=p,K,\pi$) is calculated combining measurements of the energy deposited in the MDC ($\dedx$) and the flight time in the TOF for each hadron $h$ hypothesis. 
The charged tracks are identified as pions when $\mathcal{L}(\pi)>\mathcal{L}(K)$ and $\mathcal{L}(\pi)>0$, and identified as kaons when $\mathcal{L}(K)>\mathcal{L}(\pi)$ and $\mathcal{L}(K)>0$.

The undetected neutron candidates are identified with the kinematic variable, $M_{\mathrm{miss}}^{2} \equiv E_{\rm miss}^{2}/c^{4}-\,|\vec{p}_{\rm miss}\,|^{2}/c^{2}$. Here, $E_{\rm miss}$ and $\vec{p}_{\rm miss}$ are calculated by $E_{\rm miss}\equiv E_{\rm beam}-E_{\rm rec}$ and $\vec{p}_{\rm miss}\equiv\vec{p}_{\lambdacp}-\vec{p}_{\rm rec}$, respectively, where $E_{\rm rec}~(\vec{p}_{\rm rec})$ is the energy (momentum) of the three reconstructed tracks in the $e^+e^-$ c.m. system, and the momentum of the particles from $\Ks$ decay are calculated with respect to the $\Ks$ vertex. The $\lambdacp$ momentum $\vec{p}_{\lambdacp}$ is derived by $\vec{p}_{\lambdacp}\equiv-\hat{p}_{\rm tag}\sqrt{E_{\rm beam}^{2}/c^{2}-m_{\lambdacp}^{2}c^{2}}$, where $\hat{p}_{\rm tag}$ is the momentum direction of $\lambdacm$ and $m_{\lambdacp}$ is the nominal mass of the $\lambdacp$~\cite{pdg2022}. 
The $\miss2$ spectrum is expected to peak around $0.883\;\mathrm{GeV}^{2}/c^{4}$, which is the nominal neutron mass squared~\cite{pdg2022}. 

A study of the inclusive MC sample shows that there are potential peaking backgrounds for both signal processes. For $\sigmode{1}$, we require the invariant mass differences $M_{n\pip}-M_{\rm miss}$ and $M_{n\pim}-M_{\rm miss}$ not to fall in the ranges $(0.235, 0.265)\,\gevcc$ or $(0.240, 0.270)\,\gevcc$ to eliminate the contributions from $\lambdacp\to\Sigma^{+}\pip\pim$ and $\lambdacp\to\Sigma^{-}\pip\pip$, respectively, where $M_{n\pip}$ is the invariant mass of the missing neutron and the $\pip$.
For $\sigmode{2}$, we require the invariant mass differences $M_{n\pip}-M_{\rm miss}$ and $M_{n\pim}-M_{\rm miss}$ to be outside the intervals $(0.240, 0.260)\,\gevcc$ and $(0.248, 0.268)\,\gevcc$, respectively, to suppress contributions from $\lambdacp\to\Sigma^{+}\pim K^{+}$ and $\lambdacp\to\Sigma^{-}\pip K^{+}$.

\section{\boldmath Absolute BF measurements}
\label{sec:bf}
\begin{figure*}[thp]\centering

    \subfigure[Projections of the 2D simultaneous fits to $\sigmode{1}$.]{
        \includegraphics[width=0.95\textwidth]{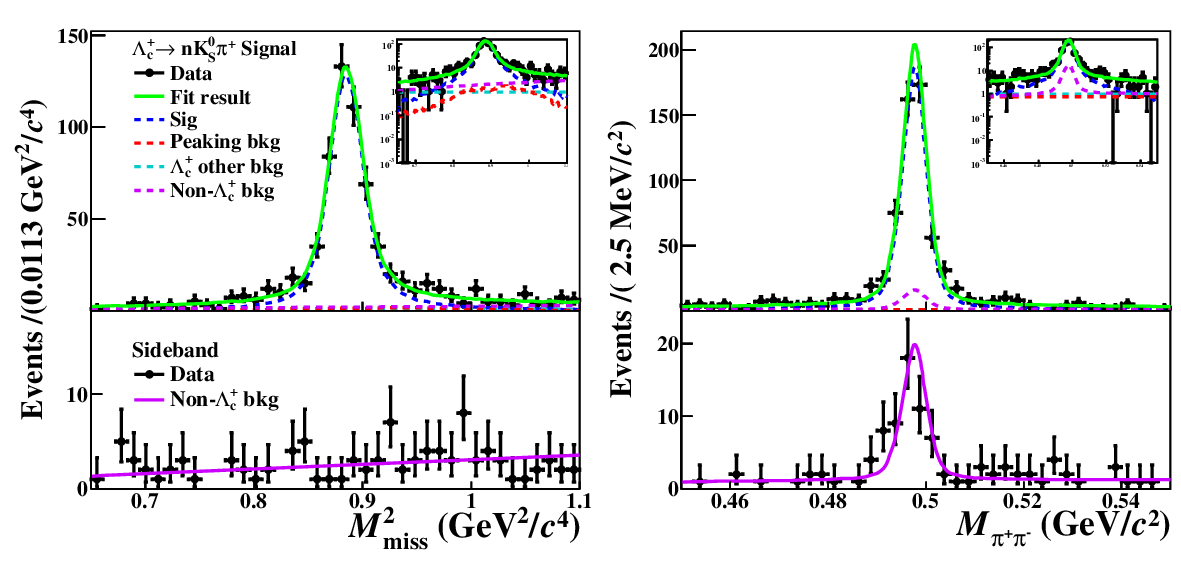}
	}
    \subfigure[Projections of the 2D simultaneous fits to $\sigmode{2}$.]{
        \includegraphics[width=0.95\textwidth]{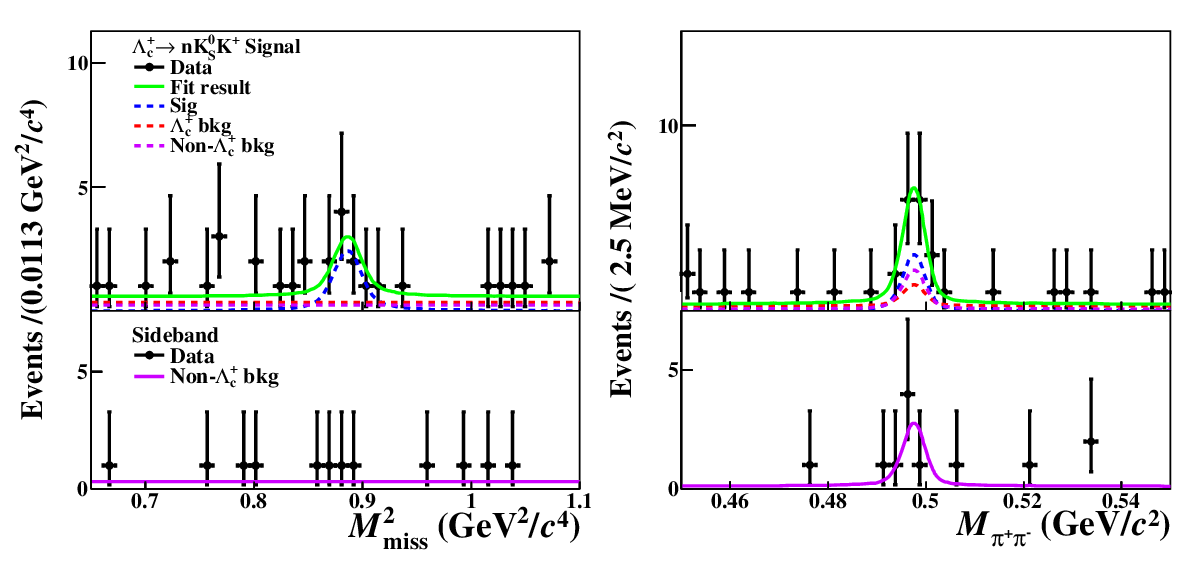}
    }
    \caption{The top and bottom sections in each of the four plots show data from the signal and sideband regions in $M_{\rm BC}$, respectively. The black dots with error bars represent data, the green solid lines represent the total fit results, the blue dashed lines represent the signal shapes, and the magenta dashed (solid) lines represent the non-$\lambdacp$ backgrounds in signal (sideband) region. For $\sigmode{1}$, the peaking and flat backgrounds from $\lambdacp$ decays are represented by the red and teal dashed lines, respectively. For $\sigmode{2}$, the $\lambdacp$ backgrounds are represented by the red dashed lines. All these signal and background components are drawn separately.}
    \label{fig:fits}
\end{figure*}

The signal yield of $\sigmode{1}$ or $\sigmode{2}$ is determined by a two-dimensional (2D) unbinned maximum likelihood fit to the spectra of $M_{\rm miss}^{2}$ and $M_{\pip\pim}$ with the combined data sets of seven energy points.
\figurename~\ref{fig:fits} shows the projections of 2D fits to the data samples for each mode. 
The signal shapes are extracted from signal MC samples and then convolved with 2D Gaussian functions accounting for the data-MC difference in the detection resolution. The parameters of Gaussian functions are derived from one-dimensional fit to the data sample of $\sigmode{1}$ and fixed in the 2D fit. 
 
For $\sigmode{1}$, a small amount of background remains which peaks in $M_{\rm miss}^{2}$ spectrum including $\lambdacp\to n\pip\pim\pip$, $\lambdacp\to\Sigma^{+}\pim\pip$, and $\lambdacp\to\Sigma^{-}\pip\pip$, which are flat in the $M_{\pi^+\pi^-}$ spectrum.

The contamination rates of these channels are estimated using exclusive background MC samples with the corresponding BFs taken from Refs.~\cite{pdg2022,BESIII:2022xne}. 
The background yields obtained are $9.2\pm0.5$, $12.7\pm1.5$ and $7.7\pm1.4$ for $\lambdacp\to n\pip\pim\pip$, $\lambdacp\to\Sigma^{+}\pim\pip$, and $\lambdacp\to\Sigma^{-}\pip\pip$, respectively. Backgrounds from other $\lambdacp$ channels are flat in both the $M_{\rm miss}^{2}$ and $M_{\pip\pim}$ spectra, which are described by a product of two flat functions in the $M_{\rm miss}^{2}$ and $M_{\pip\pim}$ dimensions.

For $\sigmode{2}$, the background processes, such as $\lambdacp\to p\Ks\piz\piz$, peak in $M_{\pip\pim}$ and are flat in $M_{\rm miss}^{2}$. Other $\lambdacp$ backgrounds are flat in both $M_{\rm miss}^{2}$ and $M_{\pip\pim}$ spectra. Therefore, the background from $\lambdacp$ decays, $f_{\lambdacp \rm bkg}$, is modeled as a product of flat background shape in the $M_{\rm miss}^{2}$ dimension and a sum of two probability density functions (PDFs) in the $\mpipi$ dimension, specifically, a constant function, $k_{0}$, and a $\Ks$ shape, $f_{\Ks~\rm shape}$, convolved with a Gaussian function, $f_{\rm Gaus}$
\begin{equation}
	f_{\lambdacp \rm bkg} \propto k_{0}\cdot\left[(1-F)\cdot k_{0}+F\cdot f_{\Ks~\rm shape}\otimes f_{\rm Gaus}\right],
	\label{eq:lcbkg}
\end{equation} 
where $F$ denotes the fraction of the $\Ks$ component which is floating in the fit.

The background originating from mis-tagged $\lambdacm$ is denoted as non-$\lambdacp$ background. Many $\Ks$ are produced in the continuum hadron process, and so the non-$\lambdacp$ background includes a peak in $M_{\pip\pim}$ and is flat in $M_{\rm miss}^{2}$. The non-$\lambdacp$ background PDF, $f_{{\rm non}\operatorname{-}\lambdacp}$, is described by a product of a polynomial, $f_{\rm Poly}$, in the $M_{\rm miss}^{2}$ dimension and a two-component PDF in the $\mpipi$ dimension. This PDF is the sum of a linear function and a $\Ks$ peak shape convolved with a Gaussian
\begin{equation}
	f_{\operatorname{non-\lambdacp}} \propto f_{\rm Poly}\cdot\left[(1-F_{2})\cdot f_{\rm Poly}+F_{2}\cdot f_{\Ks~\rm shape}\otimes f_{\rm Gaus}\right],
	\label{eq:qqbar}
\end{equation} 
 where $f_{\rm Poly}$ represents a first-order Chebyshev polynomial for $\sigmode{1}$, flat mass-independent function for $\sigmode{2}$, $F_{2}$ denotes the fraction of $\Ks$ component floated in the fit.
The yield and shape of the non-$\lambdacp$ background are shared with the data sets in the sideband region of $M_{\rm BC}$ in the ST side. 
The yield ratio between signal region and sideband region, denoted as $A$, is fixed to $1.262\pm0.005$ according to the fit to the ST $M_{\rm BC}$ distributions. 

The signal yields for $\sigmode{1}$ and $\sigmode{2}$ are $556.4\pm25.5$ and $8.8^{+3.9}_{-3.1}$, respectively, where the uncertainties are statistical. The statistical significances of $\sigmode{1}$ and $\sigmode{2}$ are $>$$10\sigma$ and $3.8\sigma$, respectively, as calculated based on the difference of the log likelihood with and without including the signal component in the fit. 

The BFs of $\sigmode{1}$ and $\sigmode{2}$ are determined by 
\begin{equation}
	\mathcal{B} = \frac{N^{\rm DT}}{\sum_{ij}N_{ij}^{\rm ST}\cdot(\varepsilon_{ij}^{\rm DT}/\varepsilon_{ij}^{\rm ST})\cdot\mathcal{B}_{\rm int}},
	\label{eq:br}
\end{equation}
where the indices $i$ and $j$ denote the ST modes and seven c.m.~energies, respectively; $\mathcal{B}_{\rm int}$ denotes the BF of $\Ks\to\pip\pim$~\cite{pdg2022}; $N^{\rm DT}$ represents the total signal yields summing over eleven ST modes and seven energy points; $\varepsilon_{ij}^{\rm DT}$, $N^{\rm ST}_{ij}$, and $\varepsilon_{ij}^{\rm ST}$ represent DT efficiencies, ST yields, and ST efficiencies, respectively. 

The determinations of the ST yields and ST efficiencies are the same as Ref.~\cite{BESIII:2022xne}. 
For $\sigmode{1}$, the hep\_ml \cite{reweightPack} package is utilized to re-weight the PHSP MC sample to consider potential intermediate states. The PHSP MC samples are trained with background-subtracted data in the $M_{n\Ks}$, $M_{n\pip}$, and $M_{\Ks\pip}$ spectra based on boosted decision trees (BDTs), and the weight is calculated accordingly for each event. The weighted signal MC samples are used to estimate the DT efficiencies for $\sigmode{1}$. The DT efficiencies for $\sigmode{2}$ are estimated by PHSP MC samples.
The ST yields, ST efficiencies, and DT efficiencies at the seven energy points are listed in TABLES~\ref{tab:N_ST}, \ref{tab:E_ST}, \ref{tab:E_DT_1}, and \ref{tab:E_DT_2}.
Finally, the BFs are determined to be $(1.86\pm0.08)\times10^{-2}$ and $\,(3.9^{+1.7}_{-1.4}\,)\times10^{-4}$ for $\sigmode{1}$ and $\sigmode{2}$, respectively, where only statistical uncertainty is considered.

\begin{table*}[htbp]
	\centering
	\caption{The ST yields, $N_{i}^{\rm ST}$, at the seven energy points and the totals. The uncertainties are statistical only.}
	\label{tab:N_ST}
	\bgroup
	\def\arraystretch{1.3}
    \resizebox{\textwidth}{!}{
    \begin{tabular}{l|cccccccc}
	   \hline\hline
	   $N_{i}^{\rm ST}$            & 4599.53 MeV    & 4611.86 MeV   & 4628.00 MeV    & 4640.91 MeV     & 4661.24 MeV   & 4681.92 MeV     & 4698.82 MeV & Total \\
       \hline
        $\bar{p}\Ks$							& $1243\pm35$ & $226\pm15$ & $994\pm33$ & $1048\pm34$ & $1044\pm33$ & $3141\pm57$ & $889\pm30$ & $8585\pm95$\\
        $\bar{p}K^{+}\pim$					& $6607\pm89$ & $1094\pm37$ & $5513\pm37$ & $5842\pm83$ & $5447\pm79$ & $15919\pm134$ & $4680\pm73$ & $45102\pm217$\\
        $\bar{p}\Ks\piz$						& $587\pm33$ & $119\pm16$ & $569\pm33$ & $552\pm33$ & $527\pm32$ & $1591\pm56$ & $414\pm30$ & $4359\pm93$\\
        $\bar{p}\Ks\pim\pip$					& $594\pm33$ & $100\pm15$ & $475\pm30$ & $484\pm30$ & $487\pm21$ & $1365\pm51$ & $414\pm28$ & $3919\pm83$\\
        $\bar{p}K^{+}\pim\piz$				& $1965\pm71$ & $331\pm30$ & $1453\pm75$ & $1458\pm63$ & $1460\pm63$ & $4361\pm109$ & $1172\pm62$ & $12200\pm188$\\
        $\bar{\Lambda}\pim$					& $738\pm27$ & $116\pm11$ & $636\pm27$ & $664\pm27$ & $624\pm26$ & $1916\pm45$ & $495\pm23$ & $5189\pm74$\\
        $\bar{\Lambda}\pim\piz$ 		        & $1681\pm54$ & $281\pm22$ & $1342\pm50$ & $1483\pm50$ & $1338\pm46$ & $3900\pm78$ & $1145\pm43$ & $11170\pm136$\\
        $\bar{\Lambda}\pim\pip\pim$			& $744\pm35$ & $130\pm14$ & $547\pm31$ & $690\pm34$ & $703\pm33$ & $1847\pm55$ & $569\pm31$ & $5230\pm93$\\
        $\bar{\Sigma}^{0}\pim$				& $502\pm25$ & $95\pm12$ & $384\pm22$ & $413\pm23$ & $414\pm22$ & $1267\pm38$ & $334\pm20$ & $3409\pm64$\\
        $\bar{\Sigma}^{-}\piz$               & $309\pm24$ & $68\pm10$ & $242\pm21$ & $271\pm22$ & $264\pm22$ & $770\pm38$ & $216\pm21$ & $2140\pm63$\\
        $\bar{\Sigma}^{-}\pim\pip$	        & $1146\pm47$ & $204\pm21$ & $922\pm19$ & $995\pm46$ & $949\pm44$ & $2729\pm79$ & $848\pm42$ & $7793\pm123$\\
        \hline
        Total &  $16116\pm157$ & $2764\pm67$ & $13077\pm125$ & $13900\pm147$ & $13257\pm140$ & $38806\pm243$ & $11176\pm133$ & $109096\pm403$\\
	   \hline\hline
   	\end{tabular}
    }
   	\egroup
\end{table*}
\begin{table*}[htbp]
	\centering
	\caption{The ST efficiencies, $\varepsilon_{i}^{\rm ST}$, at the seven energy points. The uncertainties are statistical only and the quoted efficiencies do not include the $\Ks$ BF.}
	\label{tab:E_ST}
	\bgroup
	\def\arraystretch{1.3}
    \begin{tabular}{l|cccccccc}
	   \hline\hline
	   $\varepsilon_{i}^{\rm ST} (\%)$ & 4599.53 MeV    & 4611.86 MeV   & 4628.00 MeV    & 4640.91 MeV     & 4661.24 MeV   & 4681.92 MeV     & 4698.82 MeV \\
	   \hline
	   $\bar{p}\Ks$                    & $54.6\pm0.2$ & $50.8\pm0.6$ & $48.9\pm0.2$ & $47.9\pm0.2$ & $46.4\pm0.2$ & $45.2\pm0.1$ &  $44.1\pm0.2$  \\
	   $\bar{p}K^{+}\pim$              & $49.9\pm0.1$ & $47.8\pm0.2$ & $46.1\pm0.1$ & $45.3\pm0.1$ & $44.3\pm0.1$ & $42.8\pm0.1$ &  $41.9\pm0.1$  \\
	   $\bar{p}\Ks\piz$                & $22.2\pm0.2$ & $20.8\pm0.4$ & $19.2\pm0.2$ & $19.1\pm0.2$ & $18.2\pm0.2$ & $17.6\pm0.1$ &  $16.7\pm0.2$  \\
	   $\bar{p}\Ks\pim\pip$            & $22.8\pm0.2$ & $20.4\pm0.4$ & $19.2\pm0.2$ & $19.3\pm0.2$ & $18.3\pm0.2$ & $18.7\pm0.1$ &  $17.4\pm0.2$  \\
	   $\bar{p}K^{+}\pim\piz$          & $19.4\pm0.1$ & $18.1\pm0.2$ & $16.8\pm0.1$ & $16.2\pm0.1$ & $ 15.7\pm0.1$& $15.4\pm0.0$ &  $14.9\pm0.1$  \\
	   $\bar{\Lambda}\pim$             & $47.1\pm0.3$ & $44.2\pm0.6$ & $40.7\pm0.3$ & $40.2\pm0.3$ & $38.8\pm0.3$ & $38.2\pm0.2$ &  $36.2\pm0.3$  \\
	   $\bar{\Lambda}\pim\piz$         & $20.8\pm0.1$ & $18.4\pm0.2$ & $17.6\pm0.1$ & $17.5\pm0.1$ & $ 16.9\pm0.1$& $16.1\pm0.1$ &  $15.7\pm0.1$  \\
	   $\bar{\Lambda}\pim\pip\pim$     & $15.1\pm0.1$ & $12.7\pm0.3$ & $12.7\pm0.1$ & $13.2\pm0.1$ & $12.7\pm0.1$ & $12.5\pm0.1$ &  $13.0\pm0.1$  \\
	   $\bar{\Sigma}^{0}\pim$          & $28.4\pm0.2$ & $24.8\pm0.5$ & $25.3\pm0.2$ & $24.2\pm0.2$ & $24.0\pm0.2$ & $23.2\pm0.1$ &  $21.9\pm0.2$  \\
	   $\bar{\Sigma}^{-}\piz$          & $22.8\pm0.3$ & $21.0\pm0.6$ & $21.5\pm0.3$ & $22.3\pm0.3$ & $20.5\pm0.3$ & $19.6\pm0.1$ &  $18.3\pm0.3$  \\
	   $\bar{\Sigma}^{-}\pim\pip$      & $24.5\pm0.1$ & $23.8\pm0.3$ & $21.9\pm0.1$ & $21.6\pm0.1$ & $20.9\pm0.1$ & $20.0\pm0.1$ &  $19.9\pm0.1$  \\
	   \hline\hline
   	\end{tabular}
   	\egroup
\end{table*}
\begin{table*}[htbp]
	\centering
	\caption{The DT efficiencies of $\sigmode{1}$, $\varepsilon_{i}^{\rm DT}$, at the seven energy points. The uncertainties are statistical only and the quoted efficiencies do not include the $\Ks$ BF.}
	\label{tab:E_DT_1}
	\bgroup
	\def\arraystretch{1.3}
    \begin{tabular}{l|cccccccc}
	   \hline\hline
	   $\varepsilon_{i}^{\rm DT} (\%)$ & 4599.53 MeV    & 4611.86 MeV   & 4628.00 MeV    & 4640.91 MeV     & 4661.24 MeV   & 4681.92 MeV     & 4698.82 MeV \\
	   \hline
	   $\bar{p}\Ks$                    & $23.1\pm0.1$ & $21.1\pm0.1$ & $19.7\pm0.1$& $19.2\pm0.1$& $18.9\pm0.1$& $18.5\pm0.2$ &  $17.7\pm0.1$ \\
	   $\bar{p}K^{+}\pim$              & $20.1\pm0.1$ & $19.0\pm0.1$ & $18.1\pm0.1$& $17.6\pm0.1$& $17.8\pm0.1$& $17.3\pm0.2$ &   $16.6\pm0.1$ \\
	   $\bar{p}\Ks\piz$                & $~\,8.9\pm0.1$  & $~\,8.1\pm0.1$  & $~\,7.7\pm0.1$ & $~\,7.4\pm0.1$ & $~\,7.3\pm0.1$ & $~\,7.1\pm0.1$  &  $~\,6.9\pm0.1$  \\
	   $\bar{p}\Ks\pim\pip$            & $~\,8.1\pm0.1$  & $~\,7.1\pm0.1$  & $~\,6.7\pm0.1$ & $~\,6.5\pm0.1$ & $~\,6.8\pm0.1$ & $~\,6.3\pm0.1$  &  $~\,6.0\pm0.1$  \\
	   $\bar{p}K^{+}\pim\piz$          & $~\,8.0\pm0.1$  & $~\,7.2\pm0.1$  & $~\,6.8\pm0.1$ & $~\,6.6\pm0.1$ & $~\,6.6\pm0.1$ & $~\,6.3\pm0.1$  &   $~\,6.0\pm0.1$  \\
	   $\bar{\Lambda}\pim$             & $19.5\pm0.1$ & $17.3\pm0.1$ & $16.4\pm0.1$& $16.5\pm0.1$& $15.9\pm0.1$& $15.0\pm0.2$ &  $14.5\pm0.1$ \\
	   $\bar{\Lambda}\pim\piz$         & $~\,8.3\pm0.1$  & $~\,7.3\pm0.1$  & $~\,6.9\pm0.1$ & $~\,6.6\pm0.1$ & $~\,6.5\pm0.1$ & $~\,6.4\pm0.1$  &   $~\,6.0\pm0.1$  \\
	   $\bar{\Lambda}\pim\pip\pim$     & $~\,5.5\pm0.1$  & $~\,4.8\pm0.1$  & $~\,4.6\pm0.1$ & $~\,4.7\pm0.1$ & $~\,4.6\pm0.1$ & $~\,4.4\pm0.1$  &  $~\,4.4\pm0.1$  \\
	   $\bar{\Sigma}^{0}\pim$          & $11.7\pm0.1$ & $10.5\pm0.1$ & $~\,9.8\pm0.1$ & $~\,9.9\pm0.1$ & $~\,9.3\pm0.1$ & $~\,9.2\pm0.1$  &  $~\,8.8\pm0.1$  \\
	   $\bar{\Sigma}^{-}\piz$          & $~\,9.8\pm0.1$  & $~\,9.5\pm0.1$  & $~\,8.7\pm0.1$ & $~\,8.3\pm0.1$ & $~\,8.0\pm0.1$ & $~\,7.8\pm0.1$  &  $~\,7.6\pm0.1$  \\
	   $\bar{\Sigma}^{-}\pim\pip$      & $10.0\pm0.1$ & $~\,9.1\pm0.1$  & $~\,8.8\pm0.1$ & $~\,8.6\pm0.1$ & $~\,8.5\pm0.1$ & $~\,7.9\pm0.1$  &  $~\,7.8\pm0.1$  \\
	   \hline\hline
   	\end{tabular}
   	\egroup
\end{table*}
\begin{table*}[!htbp]
	\centering
	\caption{The DT efficiencies of $\sigmode{2}$, $\varepsilon_{i}^{'\rm DT}$, at the seven energy points. The uncertainties are statistical only and the quoted efficiencies do not include the $\Ks$ BF.}
	\label{tab:E_DT_2}
	\bgroup
	\def\arraystretch{1.3}
    \begin{tabular}{l|cccccccc}
	   \hline\hline
	   $\varepsilon_{i}^{'\rm DT} (\%)$ & 4599.53 MeV    & 4611.86 MeV   & 4628.00 MeV    & 4640.91 MeV     & 4661.24 MeV   & 4681.92 MeV     & 4698.82 MeV \\
	   \hline
	   $\bar{p}\Ks$                    & $17.5\pm0.1$& $15.6\pm0.1$ & $14.9\pm0.1$& $14.6\pm0.1$& $14.4\pm0.1$ & $13.9\pm0.1$& $13.3\pm0.1$ \\
	   $\bar{p}K^{+}\pim$              & $15.6\pm0.1$& $14.5\pm0.1$ & $13.8\pm0.1$& $13.6\pm0.1$& $13.3\pm0.1$ & $13.1\pm0.1$& $12.7\pm0.1$ \\
	   $\bar{p}\Ks\piz$                & $~\,6.8\pm0.1$ & $~\,6.1\pm0.1$  & $~\,5.7\pm0.1$ & $~\,5.6\pm0.1$ & $~\,5.4\pm0.1$  & $~\,5.4\pm0.1$ & $~\,5.4\pm0.1$  \\
	   $\bar{p}\Ks\pim\pip$            & $~\,5.9\pm0.1$ & $~\,5.1\pm0.1$  & $~\,4.8\pm0.1$ & $~\,4.7\pm0.1$ & $~\,4.8\pm0.1$  & $~\,4.8\pm0.1$ & $~\,4.7\pm0.1$  \\
	   $\bar{p}K^{+}\pim\piz$          & $~\,5.6\pm0.1$ & $~\,5.3\pm0.1$  & $~\,4.9\pm0.1$ & $~\,4.8\pm0.1$ & $~\,4.5\pm0.1$  & $~\,4.5\pm0.1$ & $~\,4.4\pm0.1$  \\
	   $\bar{\Lambda}\pim$             & $14.8\pm0.1$& $13.1\pm0.1$ & $12.3\pm0.1$& $12.2\pm0.1$& $11.9\pm0.1$ & $11.7\pm0.1$& $11.1\pm0.1$ \\
	   $\bar{\Lambda}\pim\piz$         & $~\,6.2\pm0.1$ & $~\,5.4\pm0.1$  & $~\,5.1\pm0.1$ & $~\,4.9\pm0.1$ & $~\,4.7\pm0.1$  & $~\,4.6\pm0.1$ & $~\,4.6\pm0.1$  \\
	   $\bar{\Lambda}\pim\pip\pim$     & $~\,4.1\pm0.1$ & $~\,3.5\pm0.1$  & $~\,3.4\pm0.1$ & $~\,3.4\pm0.1$ & $~\,3.4\pm0.1$  & $~\,3.3\pm0.1$ & $~\,3.2\pm0.1$  \\
	   $\bar{\Sigma}^{0}\pim$          & $~\,8.9\pm0.1$ & $~\,7.9\pm0.1$  & $~\,7.5\pm0.1$ & $~\,7.3\pm0.1$ & $~\,7.0\pm0.1$  & $~\,6.9\pm0.1$ & $~\,6.7\pm0.1$  \\
	   $\bar{\Sigma}^{-}\piz$          & $~\,7.4\pm0.1$ & $~\,7.0\pm0.1$  & $~\,6.6\pm0.1$ & $~\,6.3\pm0.1$ & $~\,6.4\pm0.1$  & $~\,5.9\pm0.1$ & $~\,5.6\pm0.1$  \\
	   $\bar{\Sigma}^{-}\pim\pip$      & $~\,7.8\pm0.1$ & $~\,6.9\pm0.1$  & $~\,6.7\pm0.1$ & $~\,6.3\pm0.1$ & $~\,6.1\pm0.1$  & $~\,6.1\pm0.1$ & $~\,5.8\pm0.1$  \\
	   \hline\hline
   	\end{tabular}
   	\egroup
\end{table*}

\section{\boldmath Systematic uncertainties}
Most systematic uncertainties from the ST side cancel out in the BF measurements, as illustrated in Eq.~\eqref{eq:br}. Systematic uncertainties from different sources are summarized in \tablename~\ref{tab:syst summary} and discussed in the following.

\begin{table}[!h]
	\centering
	\caption{Relative systematic uncertainties in percentage for the BF measurements. The total systematic uncertainty is the sum in quadrature of the individual components. ``-----'' indicates cases with no uncertainty or negligible. \label{tab:syst summary}}
	\begin{tabular}{l|ccc}
	\hline\hline
	Source
		& $n\Ks\pip$ & $n\Ks K^{+}$ \\ 
	\hline
    \multicolumn{3}{c}{Multiplicative systematic uncertainties}\\
    \hline
	Extra charged track veto
		& 1.5 & 1.5 \\
	$\pi^+$, $K^+$ tracking and PID
		& 0.6 & 1.4 \\
	$\Ks$ reconstruction
		& 1.1 & 1.8 \\
	$\Sigma\pi\pi$, $\Sigma K\pi$ vetoes
		& ----- & ----- \\
	$\Sigma\pi\pi$ and $n\pip\pim\pip$ background
		& ----- & -----\\
	Intermediate BF
		& 0.1 & 0.1 \\
	MC statistics
		& 0.4 &0.4 \\
	ST fitting models
		& 0.2 & 0.2 \\
	MC model
		& 0.9 & 6.7 \\
    Multiplicative total 
		& 2.2 & 7.2\\
    \hline
    \multicolumn{3}{c}{Additive systematic uncertainties}\\
    \hline
	{Ratio $A$}
		& ----- & ----- \\
	{Fitting models in the signal side}
		&0.5 &3.4\\
    Additive total 
		&0.5 &3.4\\
	\hline
	Total
		&2.3& 8.0\\
	\hline\hline
	\end{tabular}
\end{table}

(\romanOne) \emph{No extra charged track}.
We use the control sample $\lambdacp\to\Sigma^{-}\,(n\pim)\pip\pip$ to estimate the systematic uncertainty due to the extra charged track veto. We require only three charged tracks remain recoiling against the ST side; the efficiency difference 1.5\% between data and MC simulation is assigned as the systematic uncertainty.  

(\romanTwo) \emph{$\pi^+$, $K^+$ tracking and PID}.
We select a series of control samples, $e^{+} e^{-} \rightarrow K^{+} K^{-} \pi^{+} \pi^{-}, K^{+} K^{-} K^{+} K^{-}$, $K^{+} K^{-} \pi^{+} \pi^{-} \pi^{0}, \pi^{+} \pi^{-} \pi^{+} \pi^{-}$, and $\pi^{+} \pi^{-} \pi^{+} \pi^{-} \pi^{0}$~\cite{PID_Tracking}, to study $\pi^+$, $K^+$ tracking and PID efficiencies. The momentum weighted efficiency difference between data and MC simulation is taken as the systematic uncertainty, following the method described in Refs.~\cite{BESIII:2015bjk, BESIII:2022wxj}. The combined systematic uncertainties of tracking and PID for $\pip$, $K^+$ are evaluated to be 0.6\% and 1.4\%, respectively.

(\romanThree) \emph{$\Ks$ reconstruction}.
We use the control samples $J/\psi\to K^{*}(892)^{\mp}K^{\pm}$ and $J/\psi\to\phi\Ks K^{\mp}\pi^{\pm}$. The systematic uncertainties of $\Ks$ reconstruction are 1.1\% and 1.8\% for $\sigmode{1}$ and $\sigmode{2}$, respectively.

(\romanFour) \emph{$\Sigma\pi\pi$, $\Sigma K\pi$ veto}.
We use the control samples $\lambdacp\to\Sigma^{+}\pip\pim$ and $\lambdacp\to\Sigma^{-}\pip\pip$ to study the resolution difference between data and MC simulation in the $M_{n\pip}-M_{n}$ and $M_{n\pim}-M_{n}$ spectra. The resolution difference is described by a Gaussian function which is used to correct the mass spectrum of the signal MC sample. The relative change of efficiencies before and after applying the resolution correction is taken as the systematic uncertainty.  
The mean and standard deviation of the Gaussian function are of order $10^{-4}\,\gevcc$, so the systematic uncertainty is negligible.

(\romanFive) \emph{Estimation of $\lambdacp$ peaking backgrounds}.
This uncertainty contains two parts: the contamination rates and the input BFs of $\Sigma\pi\pi$ and $n\pip\pim\pip$. 
With the same Gaussian smearing applied as for the previous $\Sigma\pi\pi/\Sigma K\pi$ veto, the difference between data and MC simulation of these backgrounds is found to be negligible.
The uncertainties of the input BFs are propagated to the peaking background yields, which are listed in Sec.~\ref{sec:bf}. 
We vary the peaking background yields within their uncertainties in the fit, and the largest difference of signal yields is also found to be negligible.

(\romanSix) \emph{Intermediate BF.}
The propagated uncertainty of the $\Ks\to\pip\pim$ BF~\cite{pdg2022} in Eq.~\eqref{eq:br} gives a 0.1\% uncertainty on the BF of signal channels.

(\romanSeven) \emph{MC statistics.}
The statistical uncertainties of DT efficiencies, ST yields and ST efficiencies are propagated to the BFs of signal channels according to Eq.~\eqref{eq:br}, which contributes a 0.4\% uncertainty.

(\romanEight) \emph{Fitting models for the ST side.}
The systematic uncertainty due to the fitting models for the ST side, 0.2\%, is quoted from Ref.~\cite{BESIII:2022xne}.

(\romanNine) \emph{MC model.}
In the nominal analysis, the DT efficiencies for $\sigmode{1}$ are estimated by the BDT-weighted signal MC sample. The hyper parameters of the BDT include the number of trees, the learning rate, maximal depth of the trees, the minimal number of events in the leaf, and the number of folds, which are (300, 0.01, 10, 200, and 3), respectively. To estimate the systematic uncertainty from the training parameters, we use another four sets of hyper parameters (250, 0.01, 10, 200, 3), (350, 0.01, 10, 200, 3), (300, 0.01, 5, 200, 3), and (300, 0.01, 15, 200, 3) to train the signal MC samples, obtaining four alternative sets of DT efficiencies. The largest difference between the alternative and nominal DT efficiencies is assigned as the systematic uncertainty, which is 0.7\%.
To estimate the systematic uncertainty from the background-subtracted data sample used in the training, we train the signal MC sample with an alternative pseudo data set to obtain another set of DT efficiencies. The alternative one is generated by randomly sampling from the nominal data set with replacement, with the sampling rate Poisson fluctuated. The difference between the nominal and alternative efficiencies, 0.6\%, is taken as the systematic uncertainty. The total systematic uncertainty from the signal MC sample for $\sigmode{1}$ is calculated to be 0.9\%.
Given the limited statistics of $\sigmode{2}$, we generate 6 sets of signal MC samples containing the resonances $\Lambda(1520)$, $\Lambda(1670)$, $\Sigma(1660)$, $\Sigma(1750)$, $a_0(980)$, and $a_2(1320)$. Seven sets of DT efficiencies are calculated based on these resonant signal MC samples and also three-body phase space. The mean value of these efficiencies is almost the same as the nominal DT efficiency. The root mean square, 6.7\%, is taken as the systematic uncertainty.

 (\uppercase\expandafter{\romannumeral10}) \emph{Ratio $A$.}
The signal-to-sideband ratio $A$ is varied $\pm1\sigma$ and alternative signal yields are obtained; the differences from the nominal yields are negligible.

(\uppercase\expandafter{\romannumeral11}) \emph{Fitting models for the signal side.}
The systematic uncertainty from the fitting model results from signal and background shapes. We vary the smearing Gaussian parameters within the uncertainties, change the flat mass-independent function (first-order Chebyshev polynomial) to a first-order (second-order Chebyshev) polynomial. 
For $\sigmode{1}$, 7000 pseudo data sets are generated randomly, where for each pseudo data set the fitting model parameters are varied randomly. 
The pull distribution of the fitted BFs in pseudo data sets indicates a relative shift of 0.5\%, which is assigned as the systematic uncertainty.
For $\sigmode{2}$, due to the limited statistics, we vary the fitting model parameters in the fit. The largest difference of the fitted signal yields from the nominal and alternative fits, 3.4\%, is taken as the systematic uncertainty.

We add the systematic uncertainties in quadrature, and the BFs for $\sigmode{1}$ and $\sigmode{2}$ are calculated to be $(1.86\pm0.08\pm0.04)\times10^{-2}$ and $\left(3.9^{+1.7}_{-1.4}\pm0.3\right)\times10^{-4}$, respectively. Here, the first uncertainties are statistical and the second systematic.
The significance considering systematic uncertainties is calculated by smearing the likelihood curve with additive systematic uncertainties. The additive systematic uncertainties include the ratio $A$ and fitting model in the signal side, while others are multiplicative, as shown in \tablename~\ref{tab:syst summary}. The multiplicative systematic uncertainties only affect the scaling of the BFs and do not affect the significance.
Finally, the significance for $\sigmode{1}$ is greater than $10\sigma$, and the significance for $\sigmode{2}$ is $3.7\sigma$.

\section{\boldmath Summary}
Based on $\ee$ collision samples with a total integrated luminosity of 4.5 $\mbox{fb$^{-1}$}$ collected with the BESIII detector at seven energy points between $4599.53\,\mev$ and $4698.82\,\mev$, we measure the absolute BF of $\sigmode{1}$ with the precision improved by a factor of 2.8~\cite{pdg2022} and report the first evidence of $\sigmode{2}$. 
The BFs for $\sigmode{1}$ and $\sigmode{2}$ are determined to be $(1.86\pm0.08\pm0.04)\times10^{-2}$ and $\left(3.9^{+1.7}_{-1.4}\pm0.3\right)\times10^{-4}$, with a significance of $>$10$\sigma$ and 3.7$\sigma$, respectively.
\begin{table}[!htbp]\footnotesize
    \renewcommand{\arraystretch}{1.25}
	\centering
	\caption{Comparisons of the BFs of $\lambdacp\to n\Ks\pip$ and $\lambdacp\to n\Ks K^{+}$ between experimental measurements and theoretical predictions.}
	\label{tab:bfComp}
	\begin{tabular}{l|ccc}
		\hline\hline
		   & $n\bar{K}^{0}\pip~(\times10^{-2})$ & $ n\bar{K}^{0}K^{+}~(\times10^{-4})$ \\
		\hline
		Geng~\cite{Geng:2018upx} & $0.9\pm0.8$ & $~\,59\pm13$ \\ 
		\hline
		Cen~\cite{Cen:2019ims} & $1.1\pm0.1$ & $31\pm9$ \\ 
		\hline
		Previous result~\cite{BESIII:2016yrc} & $3.64\pm0.50$ & -\\
		\hline
		This work & $3.72\pm0.16\pm0.08$ & $7.8^{+3.5}_{-2.8}\pm0.6$ \\
		\hline\hline
	\end{tabular}
\end{table}
\tablename{~\ref{tab:bfComp}} shows the comparison of the experimental BFs of $\sigmode{1}$ and $\sigmode{2}$ with theoretical predictions, where we assume the BFs with 
a $\bar{K}^0$ are exactly twice those observed with a $K_S^0$. 
The theoretical predictions for these two channels are based on SU(3) flavor symmetry.
The predictions for the BF of $\lambdacp\to n\Ks\pip$ are 3-4 times smaller than the experimental result from BESIII, indicating the existence of resonance states or high-wave contributions which have not been clearly identified.  
The ratio between two isospin amplitudes $R$ is evaluated to be $0.88\pm0.05$, which indicates that $I^{(1)}$ is also dominated in the dynamics, whereas $I^{(1)}$ is negligible compared with $I^{(0)}$ in the factorization scheme~\cite{Cheng:2015iom,CHENG2022324}. Hence, the factorization scheme appears to be violated in the dynamics of $\lambdacp\to n\Ks\pip$. 
Other experimental results also reveal that the factorisation scheme is violated in describing the dynamics of hadronic decays of $\lambdacp$: the measured branching fractions of the decays $\lambdacp\to\Sigma^{0}\pip$, $\lambdacp\to\Sigma^{+}\piz$, and $\lambdacp\to\Xi^{0}K^{+}$ are at the magnitude of $10^{-2}$~\cite{pdg2022}, even though no factorization diagrams contribute in these decays.
The strong phase $\cos\delta$ is calculated to be $-0.26\pm0.03$, a higher precision result than before~\cite{BESIII:2016yrc}; this is useful experimental input for understanding final state interactions in $\lambdacp$ decays and predicting the BFs of hadronic decays (for example, with final states containing a $\Lambda$ baryon~\cite{Lu:2016ogy}).
The measured BF for $\sigmode{2}$ is 3.8$\sigma$ lower (2.4$\sigma$ lower) than predicted by Geng~\cite{Geng:2018upx} (predicted by Cen~\cite{Cen:2019ims}). Thus, more theoretical work is needed to understand the three-body decays of $\lambdacp$.

\acknowledgments
The BESIII Collaboration thanks the staff of BEPCII and the IHEP computing center for their strong support. This work is supported in part by National Key R\&D Program of China under Contracts Nos. 2020YFA0406400, 2020YFA0406300; National Natural Science Foundation of China (NSFC) under Contracts Nos. 11635010, 11735014, 11835012, 11935015, 11935016, 11935018, 11961141012, 12022510, 12025502, 12035009, 12035013, 12061131003, 12192260, 12192261, 12192262, 12192263, 12192264, 12192265, 12221005, 12225509, 12235017; the Chinese Academy of Sciences (CAS) Large-Scale Scientific Facility Program; the CAS Center for Excellence in Particle Physics (CCEPP); Joint Large-Scale Scientific Facility Funds of the NSFC and CAS under Contract No. U1832207; CAS Key Research Program of Frontier Sciences under Contracts Nos. QYZDJ-SSW-SLH003, QYZDJ-SSW-SLH040; 100 Talents Program of CAS; Fundamental Research Funds for the Central Universities, Lanzhou University, University of Chinese Academy of Sciences; The Institute of Nuclear and Particle Physics (INPAC) and Shanghai Key Laboratory for Particle Physics and Cosmology; European Union's Horizon 2020 research and innovation programme under Marie Sklodowska-Curie grant agreement under Contract No. 894790; German Research Foundation DFG under Contracts Nos. 455635585, Collaborative Research Center CRC 1044, FOR5327, GRK 2149; Istituto Nazionale di Fisica Nucleare, Italy; Ministry of Development of Turkey under Contract No. DPT2006K-120470; National Research Foundation of Korea under Contract No. NRF-2022R1A2C1092335; National Science and Technology fund of Mongolia; National Science Research and Innovation Fund (NSRF) via the Program Management Unit for Human Resources \& Institutional Development, Research and Innovation of Thailand under Contract No. B16F640076; Polish National Science Centre under Contract No. 2019/35/O/ST2/02907; The Swedish Research Council; U.S. Department of Energy under Contract No. DE-FG02-05ER41374.



\end{document}